\documentclass[NETN]{stjour}
\journalname{Network Neuroscience}

\articletype{Research}

\received{X 2018}
\accepted{Y 2018}
\published{Z 2019}
\setdoi{10.1162/NETN-00001} 

\usepackage{amsmath}
\usepackage{amssymb}

\usepackage{graphicx}
\usepackage{tikz}
\usetikzlibrary{shapes}
\tikzstyle{inhibit}=[circle,draw=blue!50,fill=blue!20,thick,
inner sep=0pt,minimum size=7mm]
\tikzstyle{excite}=[circle,draw=red!50,fill=red!20,thick,
inner sep=0pt,minimum size=7mm]
\tikzstyle{pyramid}=[regular polygon,regular polygon sides=3,draw=red!50,fill=red!20,thick,
inner sep=0.3pt,minimum size=13mm]

\newcommand{\BE}{\begin{equation}}
\newcommand{\EE}{\end{equation}}

\newcommand{\BEE}{\begin{eqnarray}}
\newcommand{\EEE}{\end{eqnarray}}

\newcommand{\vecthree}[3]

\newcommand{\Red}[1]{{\color{black} #1}}
\newcommand{\Blue}[1]{{\color{black} #1}}
\newcommand{\RedText}[1]{{\color{black} #1}}

\newcommand{\e}{{\rm e}}
\renewcommand{\d}{{\rm d}}
\newcommand{\FD}[2]{\frac{\d #1}{\d #2}}

\DeclareMathSymbol{\RSet}{\mathalpha}{AMSb}{"52}
\DeclareMathSymbol{\CSet}{\mathalpha}{AMSb}{"43}

\begin{document}

\title[Node dynamics and functional connectivity]{The role of node dynamics in shaping emergent functional connectivity patterns in the brain}



\author[Author Names]
{M. Forrester\affil{1},
J. J. Crofts\affil{2},\\
S.N. Sotiropoulos\affil{3,4},
S. Coombes\affil{1} \and R.D. O'Dea\affil{1}}

\affiliation{1}{Centre for Mathematical Medicine and Biology,
School of Mathematical Sciences,
University of Nottingham,
Nottingham,
NG7 2RD, UK}

\affiliation{2}{Department of Physics and Mathematics, School of Science and Technology, Nottingham Trent University, Nottingham NG11 8NS, UK}

\affiliation{3}{Sir Peter Mansfield Imaging Centre,
Queen's Medical Centre, University of Nottingham,
Nottingham,
NG7 2UH,
UK}

\affiliation{4}{Wellcome Centre for Integrative Neuroimaging (WIN-FMRIB), University of Oxford, Oxford, OX3 9DU UK}

\correspondingauthor{M. Forrester}{Michael.Forrester@nottingham.ac.uk}


\begin{abstract}
The contribution of structural connectivity to functional brain states remains poorly understood. We present a mathematical and computational study suited to assess the structure--function issue, treating a system of Jansen--Rit neural-mass nodes with heterogeneous structural connections estimated from diffusion MRI data provided by the Human Connectome Project. Via direct simulations we determine the similarity of functional (inferred from correlated activity between nodes) and structural connectivity matrices under variation of the parameters controlling single-node dynamics, highlighting a non-trivial structure--function relationship in regimes that support limit cycle oscillations. \Red{To determine their relationship, we firstly calculate network instabilities giving rise to oscillations, and the so-called `false bifurcations' (for which a significant qualitative change in the orbit is observed, without a change of stability) occurring beyond this onset.} We highlight that functional connectivity (FC) is inherited robustly from structure when node dynamics are poised near a Hopf bifurcation, whilst near false bifurcations, structure only weakly influences FC. Secondly, we develop a weakly-coupled oscillator description to analyse oscillatory phase-locked states and, furthermore, show how the modular structure of FC matrices can be predicted via linear stability analysis. This study thereby emphasises the substantial role that local dynamics can have in shaping large-scale functional brain states.
\end{abstract}

\section{authorsummary}
Patterns of oscillation across the brain arise because of structural connections between brain regions.  
However, the type of oscillation at a site may also play a contributory role.  We focus on an idealised model of a neural mass network, coupled using estimates of structural connections obtained via tractography on Human Connectome Project MRI data.  Using a mixture of computational and mathematical techniques we show that functional connectivity is inherited most strongly from structural connectivity when the network nodes are poised at a Hopf bifurcation.  However, beyond the onset of this oscillatory instability a phase-locked network state can undergo a \textit{false bifurcation}, and structural connectivity only weakly influences functional connectivity.  This highlights the important effect that local dynamics can have on large scale brain states.

\section{Introduction}

Driven in part by advances in non-invasive neuroimaging methods that allow characterisation of the brain's structure and function, and developments in network science, it is increasingly accepted that the understanding of brain function may be obtained from a network perspective, rather than by exclusive study of its individual sub-units.
Anatomical studies using diffusion MRI allow estimation of structural connectivity (SC) of human brains, forming the so-called human connectome \citep{sporns2011human,van2013wu} which reflects white matter tracts connecting large-scale brain regions. The graph-theoretical properties of such {large-scale} networks have been well studied, highlighting key features including small-world architecture \citep{bassett2006small,liao2017small}, hub regions and cores \citep{van2013network,oldham2018development}, rich club organisation \citep{van2011rich,betzel2016optimally}, a hierarchical-like modular structure \citep{meunier2010modular,sporns2016modular}, and economical wiring \citep{bullmore2012economy,betzel2017modular}. The emergent brain activity that this structure supports can be evaluated by functional connectivity (FC) network analyses, that describe patterns of temporal coherence in neural activity between brain regions. These highly dynamic patterns are widely believed to be significant in integrative processes underlying higher brain function \citep{van2010exploring,van2013structure} and disruptions in SC and FC networks are associated with many psychiatric and neurological diseases \citep{menon2011large,braun2015human}.

However, the relationship between the brain's anatomical structure and the neural activity that it supports remains largely unknown \citep{honey2010can,park2013structural}. In particular, the divergence between dynamic functional activity and the relatively static structural connections between populations is critical to the brain's dynamical repertoire and may hold the key to understanding brain activity in health and disease \citep{park2013structural}, though current models have not yet been able to accurately simulate the transitive states underpinning cognition \citep{petersen2015brain}. Empirical studies suggest that while a structural connection between two brain areas is typically associated with a stronger functional interaction, strong interactions can nevertheless exist in their absence \citep{honey2010can,hermundstad2014structurally}; \Red{moreover, these functional networks are transient \citep{fox2005human,hutchison2013dynamic,preti2017dynamic,liegeois2017interpreting}, motivating more recent consideration of \textit{dynamic} (rather than \RedText{time-averaged}) FC networks, which have been proposed to more accurately represent brain function. An important example of SC—FC divergence is provided by resting-state networks, such as the `default mode network' and the `core network' \citep{van2010exploring,thomas2011organization}. These networks comprise brain areas that can be strongly functionally connected at rest \citep{van2010exploring}, but can also temporally vary. Indeed, a neural 'switch' has been proposed that facilitates transitions between resting--state networks \citep{goulden2014salience} and a theoretical study by \citet{messe2014relating} estimated that non-stationarity of FC contributes to over half of observed FC variance.}

Theoretical studies deploying anatomically realistic structural networks obtained through tractography alongside neural mass models describing mean-field regional neural activity have been used to further investigate the emergence of large-scale FC patterns \citep{honey2007network,rubinov2009symbiotic,deco2013resting,ponce2015resting,messe2015closer,breakspear2017dynamic}. These findings suggest that through indirect network-level interactions, a relatively static structural network can support a wide range of FC configurations; for example showing that FC reflects underlying SC on slow time scales, but significantly less so on faster time scales \citep{honey2007network,honey2009predicting,rubinov2009symbiotic}.

\Red{In the context of mean-field models, simulated (typically time-averaged) FC has been found most strongly to resemble SC when the dynamical system describing regional activity is close to a phase transition \citep{stam2016relation}, and strong structure--function agreement is reported near Hopf bifurcations in \citet{hlinka2012using}. Similarly, analysis of the dynamical systems underpinning neural simulations have shown to be a good fit to fMRI data when the system is near to bifurcation \citep{deco2019awakening,Tewarie2018}. These results provide a possible manifestation of the so-called critical brain dynamics hypothesis \citep{shew2013functional,cocchi2017criticality}. In \citet{crofts2016structure}, both SC and FC are analysed together in a multiplex network, proposing a novel measure of multiplex structure--function clustering in order to investigate the emergence of functional connections that are distinct from the underlying structure. \citet{deco2017dynamics} consider dynamic FC, with transient FC states described as meta-stable states, and in \citet{deco2019awakening}, meta-stability of a computational model of large-scale brain network activity was used to predict which structures of the brain could be influenced to force a transition between states of wakefulness and sleep. \citet{hansen2015functional} were also able to observe dynamic transitions between states resembling resting-state networks \RedText{in a noise-driven, non-linear, mean-field model of neural activity.}}

\Red{In this paper, we adopt the mean-field neural-mass approach and present a combined computational and mathematical study, which significantly extends the related works of \citet{hlinka2012using} and \citet{crofts2016structure} to investigate how the detailed and rich dynamics of the intrinsic behaviour of neural populations, together with structural connectivity, combine to shape FC networks. Thereby, we provide a complementary investigation to many of the aforementioned studies which focus on the analysis of brain networks themselves, or those that employ statistical models, by instead investigating the relationship between network structure and the emergent dynamics of these networks.} Specifically, we consider synchrony between neural subunits whose dynamics are described by the neural mass model of \citet{jansen1995electroencephalogram}, and whose connectivity is defined by a tractography-derived structural network obtained from data in the Human Connectome Project (HCP) \citep{van2013wu}. Structure--function relations are interrogated by graph-theoretical comparison of FC and SC topology under systematic variation of model parameters associated with excitatory/inhibitory neural responses, and analysed by making use of techniques from bifurcation and weakly-coupled oscillator theory.

\section{Methods}\label{sec:methods}

\subsection{Neural mass model}\label{subsec:model}

We consider a network of interacting neural populations, representing a parcellation of the cerebral cortex, such that each area (node) corresponds to a functional unit that can be represented by a neural mass model, and with edges informed by structural connectivity. \Red{Neural mass activity is represented by the Jansen--Rit model \citep{jansen1995electroencephalogram} of dimension $m=6$, that describes the evolution of the average post-synaptic potential (PSP) in three interacting neural populations: pyramidal cells ($y_0$), and excitatory ($y_1$) and inhibitory ($y_2$) interneurons. These populations are connected with strengths $C_i$ ($i=1...4$), representing the average number of synaptic connections between each population. The Jansen--Rit model is mathematically described by three second order ordinary differential equations which are commonly rewritten as six first order equations by adopting the notation $(y_0,\ldots,y_5)$ for the dependent variables. The pairs $(y_0,y_3)$, $(y_1,y_4)$, and $(y_2,y_5)$ are therefore associated with the dynamics of the \RedText{population average of PSPs} and their temporal derivatives. The quantity of primary interest herein is $y=y_{1}-y_{2}$, which is physiologically interpreted as the average potential of pyramidal populations and the main contributor to signals generated in EEG recordings \citep{teplan2002fundamentals}.} Introducing an index $i=1,\ldots, N$ to denote each node in a network of $N$ interacting neural populations, we write the evolution of state variables as:
\begin{align}
\label{eq:JR}
\dot{y}_{0_i}=\quad &y_{3_i}, \qquad \dot{y}_{1_i}\quad =y_{4_i}, \qquad \dot{y}_{2_i}=\quad y_{5_i}, \nonumber \\
\dot{y}_{3_i}=\quad &Aa f\left(y_{1_i}-y_{2_i}\right)-2ay_{3_i}-a^2y_{0_i},\\
\dot{y}_{4_i}=\quad &Aa\left\{P_i+\varepsilon\sum_{j=1}^N{w_{ij}f\left(y_{1_j}-y_{2_j}\right)}+C_2f\left(C_1y_{0_i}\right)\right\} -2ay_{4_i}-a^2y_{1_i}, \nonumber\\
\dot{y}_{5_i}=\quad &BbC_4f\left(C_3y_{0_i}\right)-2by_{5_i}-b^2y_{2_i} .\nonumber
\end{align}
Here $f$ is a sigmoidal nonlinearity, representing the transduction of activity into a firing rate, and with the specific form
\begin{equation}
f(v)=\frac{\nu_{\text{max}}}{1+\mbox{exp}(r(v_0-v))}.
\label{eq:sigmoid}
\end{equation}

\begin{figure}[ht]
\begin{center}
\begin{tikzpicture}[->,shorten >=1pt,auto,node distance=3.5cm,semithick]
	\node (y1) at (0,0) [excite] {$E$};
	\node (y2) at (4,0) [inhibit] {$I$};
	\node (y0) at (2,3) [pyramid] {$PC$};
	
	\path (y0) edge [bend left=10,color=red] node {\Red{$C_1$}} (y1)
	      (y0) edge [bend left=10,color=red] node {\Red{$C_3$}}  (y2)
	      (y1) edge [bend left=10,color=red] node {\Red{$C_2$}} (y0)
	      (y2) edge [bend left=10,color=blue] node {\Blue{$C_4$}} (y0);
	      
	\node [left=2cm,text width=4cm,font=\footnotesize] at (y0)
{
\Red{$\quad P_i+\varepsilon\sum\limits_jw_{ij}f(y_{1_j}-y_{2_j})$}
}
	edge [color=red] (y0);
\end{tikzpicture}
\end{center}
\caption{\Red{Wiring diagram for a Jansen-Rit network node, described by equations (1,2). Excitatory/inhibitory populations and synaptic connections are highlighted in red/blue respectively. Interneurons ($E,I$) and pyramidal cells ($PC$) are interconnected with strengths $C_i$ for $i=1...4$. \RedText{Also shown is the expression for the external input to a $PC$ population}, consisting of a \RedText{extracortical input $P_i$}, as well as contributions from afferently connected nodes.}}\label{fig:schematic}
\end{figure}
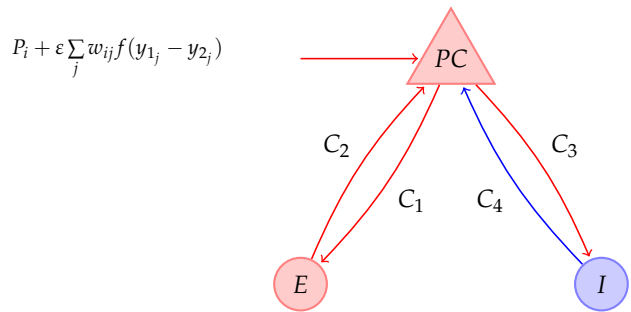

\Red{The model is identical to that presented in \citet{jansen1995electroencephalogram} for a single cortical column, but is completed by the specifying the network interactions as a function of average membrane potential of afferently connected pyramidal populations, encoded in a connectivity matrix with elements $w_{ij}$ (described in \textbf{\textit{Structural and functional connectivity}}), with an overall scale of interaction set by $\varepsilon$.} The remaining model parameters, together with their physiological interpretations \Red{and values (taken from \citet{grimbert2006analysis}, and \citet{Touboul2011})}, are given in Table \ref{tab:JR}. \Red{A schematic `wiring diagram' for the model indicating the interactions between different neural populations is shown in Fig.~\ref{fig:schematic}.}

The Jansen--Rit model, defined by equation \eqref{eq:JR}, can support oscillations that relate to important neural rhythms, such as the well known alpha, beta and gamma brain rhythms, and also irregular, epileptic-like activity \citep{ahmadizadeh2018bifurcation}. Moreover, the model is able to replicate visually-evoked potentials seen in EEG recordings \citep{jansen1995electroencephalogram}, from which FC may be empirically measured \citep{srinivasan2007eeg}. 

\begin{table}[ht!]
\centering
\small{\begin{tabular}{|>{\centering\arraybackslash}p{0.15\textwidth}|>{\centering\arraybackslash}p{0.5\textwidth}|>{\centering\arraybackslash}p{0.2\textwidth}|}
\hline 
Parameter & Meaning & Value\\ 
\hline
$C_1$, $C_2$, $C_3$, $C_4$ & Average number of synapses between populations & $135$, $108$, $33.75$, $33.75$\\ 
\hline
$P_i$ & \RedText{ Basal extracortical input to} main pyramidal excitatory populations & $120$ Hz\\ 
\hline
$A,B$ & Amplitude of excitatory, inhibitory PSPs respectively & $[2,14]$ mV, $[10,30]$ mV\\ 
\hline
$a,b$ & Lumped time constants of excitatory, inhibitory PSPs & $100\mbox{ s}^{-1}$, $50\mbox{ s}^{-1}$\\
\hline
$\varepsilon$ & Global coupling strength & 0.1\\ 
\hline
$w_{ij}$ & Coupling from node $j$ to $i$ & $[0,1]$\\ 
\hline
$\nu_{\mbox{max}}$ & Maximum population firing rate & $5$ Hz\\ 
\hline
$v_0$ & Potential at which half-maximum firing rate is achieved & $6$ mV\\ 
\hline
$r$ & Gradient of sigmoid at $v_0$ & $0.56\mbox{ mV}^{-1}$\\ 
\hline
\end{tabular} 
}
\vspace*{.5cm}
\caption {\Red{Parameters in the Jansen--Rit model, given by equations \eqref{eq:JR} and \eqref{eq:sigmoid} along with physiological interpretations and values/ranges used in simulations, which were taken from \citet{grimbert2006analysis} and \citet{Touboul2011}.} In particular, the values $A$ and $B$, which modulate the strength of excitatory and inhibitory responses respectively, were chosen as the key control parameters for varying network activity.}
\label{tab:JR}
\end{table}

In what follows, we consider the patterns of dynamic neural activity that arise under systematic variation of the model parameters $A$ and $B$, these being chosen as the parameters of interest because they govern the interplay between inhibitory and excitatory activity, which would typically vary due to neuromodulators in the brain \citep{rich2018effects}. It is known that a single Jansen--Rit node can support multi-stable behaviour which includes oscillations of different amplitude and frequency but, moreover, a network of these nodes can also exhibit various stable phase-locked states. \Red{A small amount of white noise is added to the \RedText{extracortical input $P_i$} on each node, in order to allow the system to explore a variety of these dynamical states: \RedText{$P_i+dW_i(t)$, where $dW_i(t)$ is chosen at random from a Gaussian distribution with standard deviation $10^{-1}$ Hz and mean 0 Hz.} For direct simulations of the network we use an Euler--Murayama scheme, implemented in Matlab\textregistered, with a fixed numerical time-step of $10^{-4}$, which we have confirmed ensures adequate convergence of the method.}

\subsection{Structural and functional connectivity}\label{subsec:conn}

\begin{figure}[ht]
	\centering
		\includegraphics[width=\textwidth]{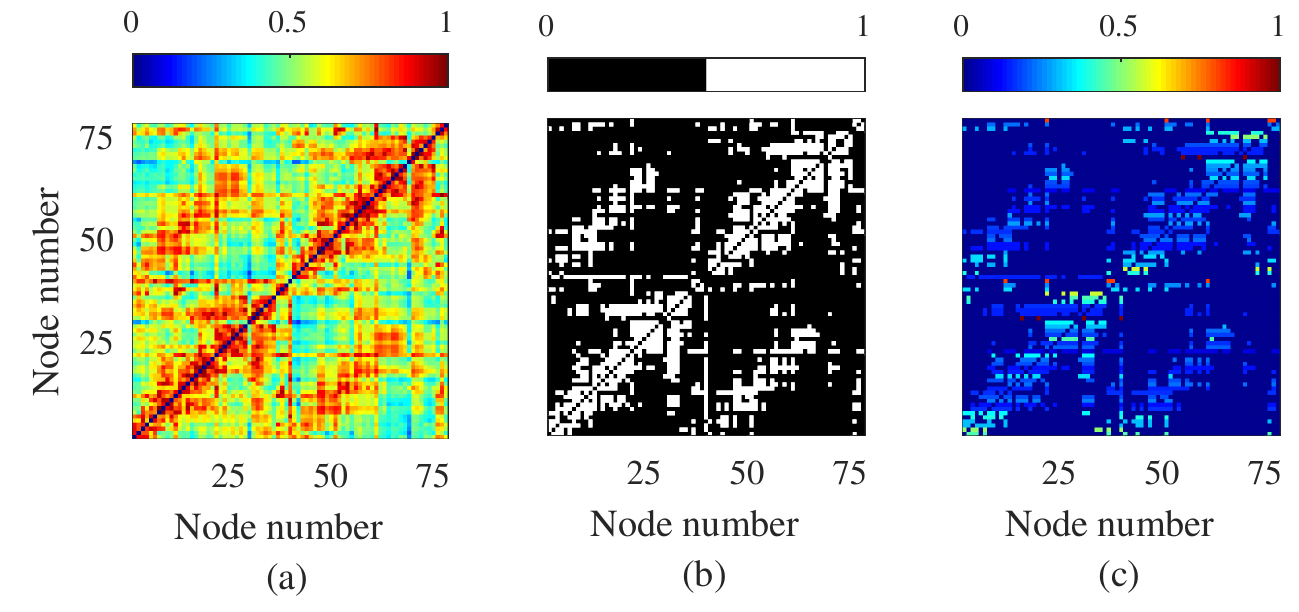}
	\caption{The original structural matrix (a) is derived from DTI data taken from the Human Connectome Project database and parcellated on to a 78-region brain atlas. This is thresholded and binarised to keep the top 23\% strongest connections (b) and normalised by row so that $\sum_{j=1}^N w_{ij}=1$ for all regions $i$) in (c).}
	\label{fig:Networks}
\end{figure}

The structural connectivity was estimated using diffusion MRI data recorded with informed consent from 10 subjects, obtained from the HCP \citep{van2013wu}. Briefly, we explain how this data is post-processed to derive connectomic data, though we direct the reader to \citet{tewarie2019spatially} and the references therein for a more detailed overview. 60,000 vertices on the white/grey matter boundary surface for each subject \citep{glasser2013minimal} were used as seeds for 10,000 tractography streamlines. Streamlines were propagated through voxels with up to three fibre orientations, estimated from distortion-corrected data with a deconvolution model \citep{jbabdi2012model,sotiropoulos2016fusion}, using the FSL package. The number of streamlines intersecting each vertex on the boundary layer was measured and normalised by the total number of valid streamlines. This resulted in a 60,000 node structural matrix, which was further parcellated using the 78-node AAL atlas. This was used to describe connections between brain regions, providing an undirected (symmetric), weighted matrix whose elements $w_{ij}$ define the strengths of the excitatory connections in equations \eqref{eq:JR}. \Red{To enable a meaningful comparison between the network measures of SC and FC, the former reflecting the density of tractography streamlines and the latter that of correlated neural activity, we place them on a similar footing by thesholding and binarising, such that only the top 23\% of the weights (ordered by strength) are retained; see Fig.~\ref{fig:Networks}.  Thresholding is a widespread technique for removing spurious connections that may not in fact be a realistic representation of brain connectivity.  We note that our thresholding choice (that reduces the number of connections, while ensuring that the overall modular structure is unchanged) is commensurate with a recent study \citep{tsai2018reproducibility}, which employed DTI data averaged on the same brain atlas as used herein to consider thresholding approaches suitable to remove weak connections with high variability between ($n=30$) different subjects. To generate nodal inputs with commensurate magnitudes, the structural connectivity matrix was normalised by row so that afferent connection strengths for each node sum to unity. This normalisation process permits some of the analysis that we undertake to help explain SC--FC relations (see \textbf{\textit{Weakly coupled oscillator theory}}); however, we highlight that the results that we present herein are not crucially dependent on such a choice and so our conclusions generalise (see \textbf{MATHEMATICAL METHODS}).}

\RedText{In view of the non-linear oscillations supported by the network model given by \eqref{eq:JR}, functional connectivity networks are obtained by computing the commonly-used metric of mean phase coherence (MPC; \citet{mormann2000mean}), which determines correlation strength in terms of the proclivity of two oscillators to phase-lock, giving a range from 0 (completely desynchronised) to 1 (phase-locking). We choose $y_j=y_{1_j}-y_{2_j}$ as the variable of interest because of its relation to the EEG signal, making it a good candidate to produce timeseries more readily comparable with empirical data. Pairwise MPC measures the average temporal variance of the phase difference $\Delta \phi_{jk}(t) = \phi_j(t)-\phi_k(t)$, between two time-series indexed by $j$ and $k$, where here the instantaneous phase $\phi_j(t)$ is obtained as the angle of the complex output resulting from application of a Hilbert transform to the time-series, $y_j(t)$. The mean phase coherence of the time-series comprising $M$ time-points $t_l$ ($l=1, \ldots, M$) is defined as:
\begin{equation}
	R_{jk} = \bigg| \frac{1}{M}\sum\limits_{l=1}^{M}\mathrm{e}^{i\Delta\phi_{jk}(t_l)} \bigg|.\label{eq:Rjk}
\end{equation}}

Structure--function relations are assessed by computing the Jaccard similarity coefficient \citep{jaccard1912distribution} of the non-diagonal entries of the binarised SC and FC matrices. This describes the relative number of shared pairwise links between the two networks, providing a natural measure of structure--function similarity, ranging from zero for matrices with no common links to unity for identical matrices.

\Red{Since the SC--FC correlation patterns of interest here arise naturally from global synchrony or patterns of phase-locking of oscillatory node activity, the \RedText{local stability} of oscillatory node dynamics and of network (global or phase-locking) synchrony is a natural candidate to explain the structures we observe.  In the following subsections we consider bifurcation, false bifurcation and weakly-coupled oscillator theory approaches to address this.}
\subsection{Bifurcation analysis}\label{subsec:bif}

\subsubsection{Single node and network bifurcations}
\label{subsec:nodebif}

\begin{figure}[ht]
\centering
		\includegraphics[width=0.6\textwidth]{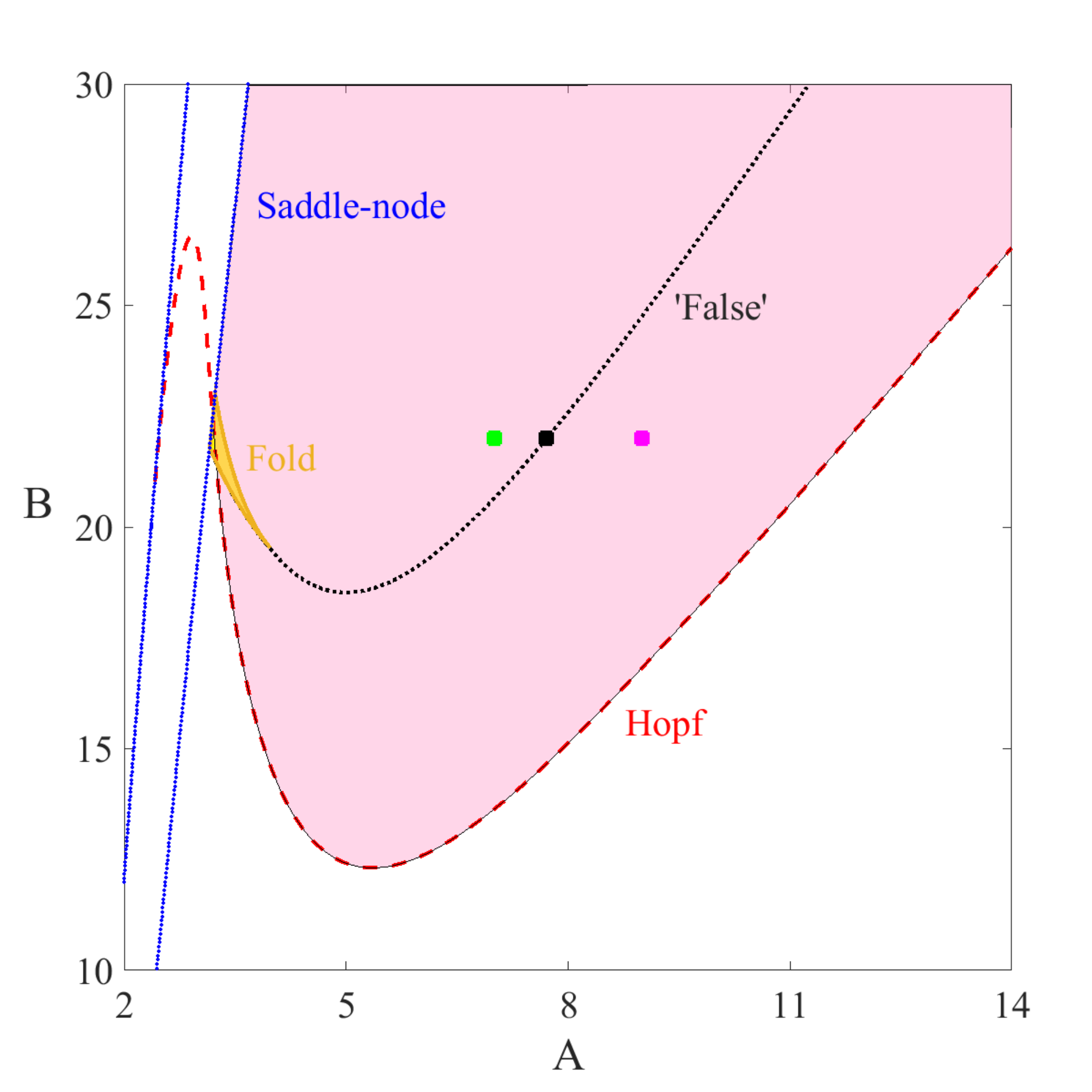}
	\caption{Two-parameter bifurcation diagram in the $(A,B)$ plane in the single-node case of the Jansen--Rit system of equations \eqref{eq:JR}. Other parameter values are as stated in Table \ref{tab:JR}. Red dashes are Hopf bifurcations, black dots are false bifurcations and blue lines represent saddle points. There is also a region of bistability, highlighted in yellow, which is bounded by saddle nodes and a set of fold bifurcations of limit cycles. The pink and yellow shaded regions indicates parameter values for which there exist stable oscillatory solutions. The three coloured dots at $B=22$, $A=7.0,\ 7.7,\ 9.0$ indicate parameter values at which we observe distinctly different dynamics as shown in Fig. \ref{fig:activity}.}
	\label{fig:bif}
\end{figure}

Bifurcations for a single node are readily computed using the software package XPPAUT \citep{ermentrout2002simulating}, using $A$ and $B$ as the parameters of interest. The result is a Hopf and saddle-node set in parameter space, which bounds a region of oscillatory solutions. \Red{We also observe a region of bistability bounded by fold bifurcations of limit cycles, in which the types of activity described in Fig.~\ref{fig:activity}(a) and (c) can both exist.} This is shown in Fig.~\ref{fig:bif}. \Red{We refer the reader to \citet{grimbert2006analysis} \citet{Touboul2011} and \citet{spiegler2010bifurcation} for a comprehensive analysis of the bifurcation structure of the Jansen--Rit model.}

The corresponding diagram for the full network requires numerical analysis of a much higher dimensional system, described by $N \times m = 78 \times 6 = 468$ ODEs; this is computationally demanding, and so in \textbf{MATHEMATICAL METHODS} we develop a quasi-analytic approach by linearising the full network equations around a fixed point. The resulting equations can be diagonalised in the basis of eigenvectors of the structural connectivity, leading to a set of $N$ equations, each of which prescribes the spectral problem for an $m$-dimensional system.  Thus, each of these low dimensional systems can be easily treated without recourse to high performance computing.  Moreover, this approach exposes the role that the eigenmodes of the structural connectivity matrix has in determining the stability of equilibria. We report the locus of Hopf and saddle-node sets for the network in Fig. \ref{fig:results}. Comparison of Figs \ref{fig:bif} and \ref{fig:results} shows that the bifurcation structure of steady states for the full network is practically identical to that of the single node (even for moderate coupling strength---here, $\varepsilon=0.1$), highlighting the potential importance of single-node dynamics in driving SC--FC correlations.

\subsubsection{False bifurcations}\label{subsec:falsebif}

\begin{figure}[ht]
\centering
	\includegraphics[width=\textwidth]{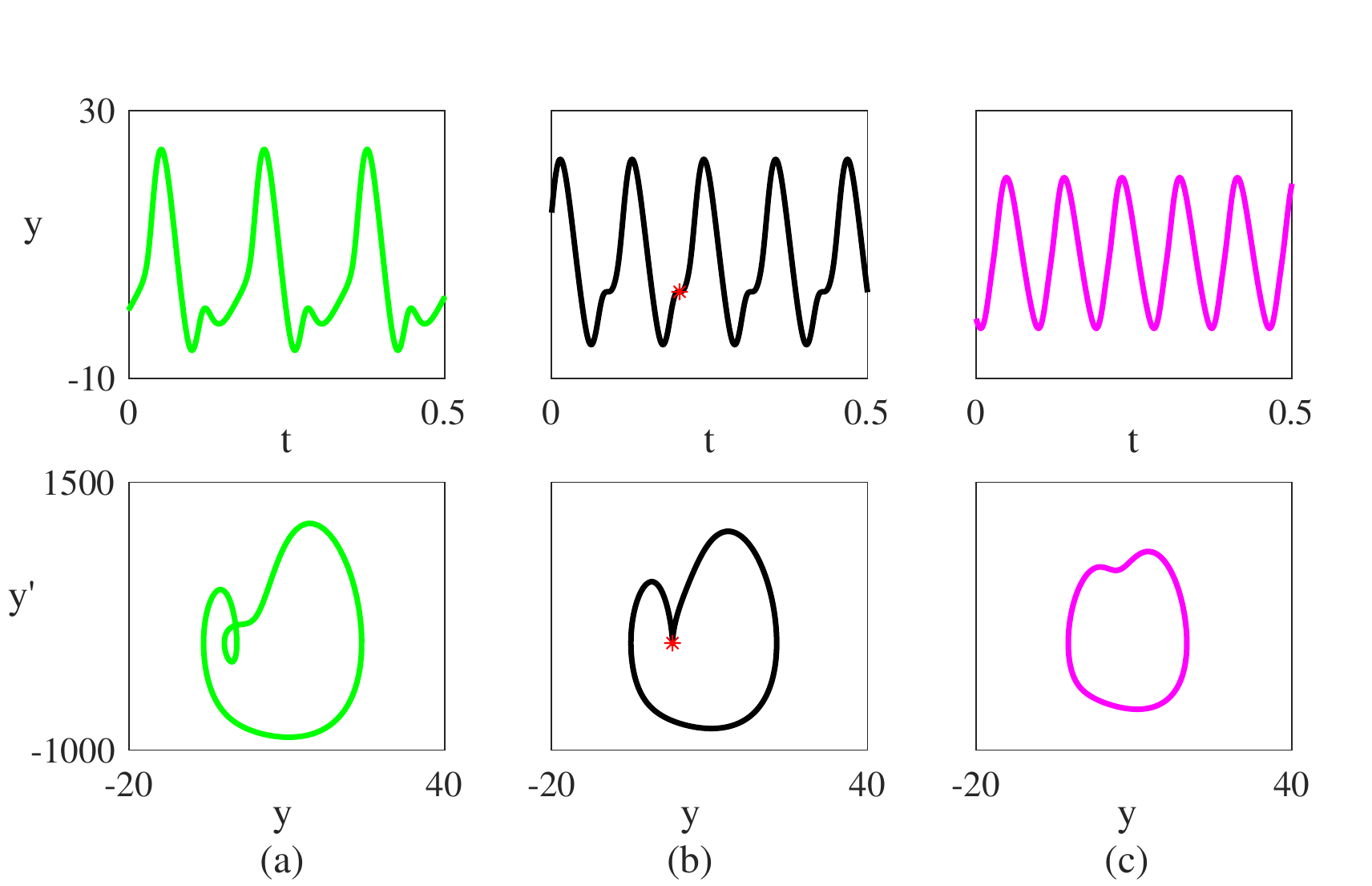}
	\caption{Activity profiles of $y=y_1-y_2$, the potential of the main population of pyramidal neurons for a node in the Jansen--Rit network \Red{(1) in the absence of noise}, with $B$ fixed at $22$ and (a) $A=9.0$; (b) $A=7.7$; (c) $A=7.0$ and other parameter values as in Table \ref{tab:JR}. Subfigures in the upper row are plots of the timeseries solution, whereas the bottom row shows the trajectories of stable orbits in the $(y,y')$ plane. The chosen parameters lie at either side of the region where a smooth transition between activity types occurs, corresponding to a \textit{false bifurcation} (see highlighted parameter values in Fig.~\ref{fig:bif}). In (b), an inflection point occurs and is highlighted as a red star on the orbit.}
\label{fig:activity}
\end{figure}

In Fig.~\ref{fig:activity} we consider in more detail the types of activity that the network model \eqref{eq:JR} supports. In particular, we observe that under changes to parameter values within the oscillatory region (see highlighted parameter values in Fig. \ref{fig:bif}), the time-course of activity shifts from single- to double-peaked waves, which could have consequences for synchronisation of oscillations and, moreover, FC. The points of transition are known as \textit{false bifurcations} since there is a significant dynamical change that occurs smoothly rather than critically. False bifurcations in a neural context have previously been seen as canards in single neuron models \citep{Desroches2013} as well as in EEG models of absence seizures \citep{Marten2009}.  In the latter case the false bifurcation corresponds to the formation of spikes associated with epileptic seizures \citep{moeller2008simultaneous}.

As illustrated in Fig.~\ref{fig:activity} the false-bifurcation transition is characterised by the change from a double-peaked profile (a) to a sinusoidal-like waveform (c) via the development of a point of inflection in the solution trajectory (b). Since this transition is not associated with a change in stability of the periodic orbit, these \textit{false bifurcations} are determined by tracking parameter sets for which points of inflection occur. We refer the reader to \citet{rodrigues2010method} for details on methods for detecting and continuing false bifurcations in dynamical systems. The result of this computation is shown in Fig.~\ref{fig:bif}, \Red{where we observe the set of false bifurcations arising from the breakdown of two branches of fold bifurcations of limit cycles. In the full network (not shown), this computation is more laborious  (and there is some delicacy in defining the bifurcation since the network coupling leads nodes to inflect at marginally different parameter values); however, we obtain very similar results to those obtained in Figure 3 for a single node (not shown).}

\subsection{Weakly-coupled oscillator theory}\label{subsec:weak}

Further insight into the phase relationship between nodes in a network can be obtained from the theory of weakly coupled oscillators (see, \textit{e.g.},~\citet{hoppensteadt2012weakly}).  This technique reduces a network of limit cycle oscillators to a set of relative phases in a systematic way.  The resulting set of network ODEs is $(N-1)$-dimensional, as opposed to the $(Nm)$-dimensionality of the original system, and provides an accurate model as long as the overall coupling strength is weak ($|\varepsilon| \ll 1$). This is because when all oscillators lie on the same limit cycle of a system, the interactions from pairwise-connected nodes can be considered as small perturbations to the oscillator dynamics. Moreover, the resulting set of network ODEs only depends upon phase differences and it is straightforward to construct relative equilibria (oscillatory network states) and determine their stability in terms of both local dynamics and structural connectivity.  A method to construct the \textit{phase interaction function}, $H$, for the network is provided in \textbf{MATHEMATICAL METHODS}.  Once this is known, the dynamics for the phases of each node in the network, $\theta_i \in [0, 2 \pi)$, takes the simple form:
\begin{equation}
	\dot\theta_i = \Omega + \varepsilon\sum_{j=1}^N w_{ij}H(\theta_j-\theta_i), \qquad i=1,\ldots,N-1,
	\label{eq:phasedynamics}
\end{equation}
where $\Omega=2\pi/T$ represents the natural frequency of an uncoupled oscillatory node with period $T$, and the second term determines phase changes arising from pairwise interactions between nodes.  We emphasise that the $T$-periodic phase interaction function $H(\Omega t) = H(\Omega(t+T))$ is \textit{derived} from the full system given by (\ref{eq:JR}). For a given phase-locked state $\theta_i(t) =\Omega t + \phi_i$ (where $\phi_i$ is the constant phase of each node), \RedText{local stability} is determined in terms of the eigenvalues of the Jacobian of \eqref{eq:phasedynamics}, denoted by $\widehat{H}(\mathbf{\Phi})$ with $\mathbf{\Phi} = (\phi_1, \ldots, \phi_N)^\intercal$, with components:
\begin{equation}
	[\widehat{H}(\mathbf{\Phi})]_{ij}=\varepsilon [H'(\phi_j-\phi_i) w_{ij} - \delta_{ij} \sum_{k=1}^{N} H'(\phi_k-\phi_i) w_{ik}].
	\label{eq:H_Jacobian}
\end{equation}
The globally synchronous steady-state, $\phi_i = \phi$ for all $i$, exists in a network with a phase interaction function that vanishes at the origin (\textit{i.e.} $H(0)=0$, which is not the case here), or for one with a row-sum constraint, $\sum_j w_{ij} = \Gamma = \text{constant}$ for all $i$, which is true for our specific structural matrix (for which $\Gamma=1$). Note that the emergent frequency of the synchronous network state is given explicitly by $\Omega+\varepsilon \Gamma H(0)$. Using the Jacobian in \eqref{eq:H_Jacobian}, synchrony is found to be stable if $\varepsilon H'(0) >0$ and all the eigenvalues of the graph Laplacian of the structural network,
\begin{equation}
	[\mathcal{L}]_{ij}=-w_{ij} + \delta_{ij} \sum_{k} w_{ik},
\end{equation}
lie in the right hand complex plane. Since the eigenvalues of a graph Laplacian all have the same sign (apart from, in this case, a single zero value) then \RedText{local stability} is entirely determined by the sign of $\varepsilon H'(0)$.  For example, for a globally coupled network with $w_{ij}=1/N$ then the graph Laplacian has one zero eigenvalue, and $(N-1)$ other degenerate eigenvalues at $-1$, and so synchrony is stable if $\varepsilon H'(0) >0$. 

It is therefore useful to consider the condition $\varepsilon H'(0) > 0$ as a natural prerequisite for a structured network to support high levels of synchrony (without recourse to exploring the full Jacobian structure). A plot of $\varepsilon H'(0)$ is shown in Fig.~\ref{fig:results}(b). For completeness, however, the full Jacobian was also computed in order to account for the potential influence of detailed structure on the correspondence with the observed SC--FC agreement measured in simulations. To do this, the system given by \eqref{eq:JR} was integrated with $\varepsilon=0.001$ to a (stable) phase-locked state, and relative phases computed.  The eigenvalues of the Jacobian (eq. \eqref{eq:H_Jacobian}) were then computed, providing an indication of solution attractivity. The largest non-zero eigenvalue for each parameter choice is shown in Fig. \ref{fig:results}(c).

It has been shown in \citet{Tewarie2018} that the eigenmodes of the structural connectivity matrix are predictive of emergent FC networks arising from an instability of a steady state. The largest non-zero eigenvalue, which is related the most unstable eigenmode (or closest to instability), was found to be a good predictor of resultant FC by computing the tensor product of its corresponding eigenvector, $v\otimes v$. Here we take this further by considering instabilities of the \textit{synchronous} state. 
In this case the Jacobian \eqref{eq:H_Jacobian} reduces to $-\varepsilon H'(0) \mathcal{L}_{ij}$ and the phase-locked state that emerges beyond instability of the synchronous state has a pattern determined by the a linear combination of eigenmodes of the graph Laplacian, since all eigenmodes destabilise simultaneously.
It is known that the graph Laplacian can be used to predict phase-locked patterns \citep{chen2012laplacian} and has indeed been used to predict empirical FC from SC \citep{abdelnour2018functional}. \Red{Following from this, the eigenmodes of the Jacobian in \eqref{eq:H_Jacobian} can be used as simple, easily computable proxy for the FC matrix when the system is poised at a \RedText{local instability.}} In Fig. \ref{fig:eig} we compare the FC pattern from the (fully nonlinear) weakly coupled network with a linear prediction, to highlight its usefulness. \RedText{In this case, MPC \eqref{eq:Rjk} is not ideally suited for our study because it struggles to discern between phase-locking and complete synchrony, yet we consider situations where stable phase-locking naturally arises. Therefore, FC in the weakly-coupled network is computed via the new metric of mean phase agreement (MPA), whereby patterns of coherence are determined by a temporal average of relative phase differences:
\begin{equation}
	\hat{R}_{jk} = \frac{1}{M}\sum\limits_{l=1}^{M}\frac{1}{2}\big(1+\cos(\Delta\phi_{jk}(t_l))\big).\label{eq:MPA}
\end{equation}}
For comparison, we use the tensor product sum,
\begin{equation}
	\hat{R} = \sum_{i=1}^{N^*} \lambda_i v_i \otimes v_i
	\label{eq:Rjk_predicted}
\end{equation}
of $v_k=(v_k^1,\ldots,v_k^N)$, which denotes the $k^{\mbox{th}}$ eigenvector of the Jacobian for the synchronous state. These are weighted by their corresponding eigenvalues, $\lambda_k$, and we include the $N^*$ unstable eigenmodes.
\section{Results}\label{sec:results}

\begin{figure}[!ht]
\centering
		\includegraphics[width=\textwidth]{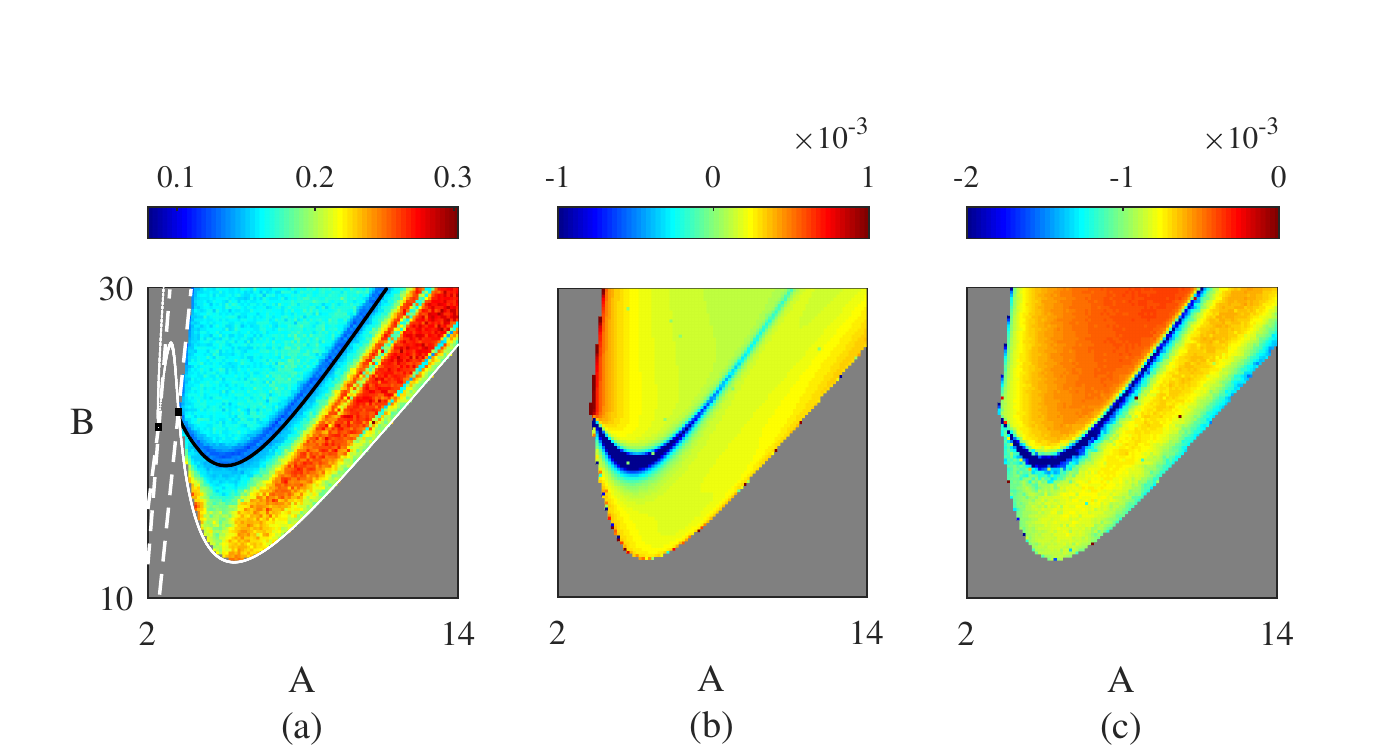}
	\caption{(a) Jaccard similarity coefficient between SC and FC \RedText{(measured by MPC in \eqref{eq:Rjk})} when the Jansen--Rit network \Red{(\ref{eq:JR})} supports an oscillatory solution, averaged over 30 realisations of initial conditions chosen at random. Parameter values are given in Table \ref{tab:JR}. Warmer colours indicate greater SC/FC correlation. Here we have superimposed the bifurcation diagram for the network steady state, which shows the oscillatory region being bounded by Hopf/saddle-node sets in solid/dashed white lines respectively; boxes are Bogdanov--Takens points. False bifurcations in the single node case are indicated by a black line \Red{but, because of its relative size, the bistable region is not shown (though can be seen for the single node case in Fig.~\ref{fig:bif}).} (b) The value of $H'(0)$ \Red{(see eqs.~(\ref{eq:phasedynamics},\ref{eq:H_Jacobian}))} in the $A,B$-plane. When this value is positive/negative, the globally synchronised solution is stable/unstable (if it exists); (c) The largest non-zero eigenvalue of the Jacobian for the full weakly-coupled oscillator network \Red{(equation \eqref{eq:H_Jacobian})}, calculated at a stable phase-locked state. More negative values indicate a stronger stability.}
\label{fig:results}
\end{figure}

Fig.~\ref{fig:results} shows plots in the $(A,B)$ parameter space highlighting our studies on the combined influence of SC and node dynamics on FC. The region bounded by the bifurcation curves, obtained via a linear instability analysis of the network steady state, is where the network model supports oscillations as well as phase-locked states.
In Fig.~\ref{fig:results}(a) the Jaccard similarity between SC and FC is computed from direct numerical simulations of the Jansen--Rit network model (\ref{eq:JR}).
Beyond the onset of oscillatory instability (supercritical Hopf bifurcation) the emergent phase-locked network states show a nontrivial correlation with the SC.  This varies in a rich way as one traverses the $(A,B)$ parameter space, showing that precise form of the node dynamics can have a substantial influence on the network state.
The highest correlation between SC and FC coincides with a Hopf bifurcation of a network equilibrium (shown as a solid white line), whilst a band of much lower correlation coincides with \Red{the fold bifurcations of limit cycles} and false bifurcations of a single node (in black), reproduced from Fig. \ref{fig:bif}. Indeed, it would appear that these mathematical constructs are natural for organising the behaviour seen in our \textit{in silico} experiments. \Red{We reiterate that we have confirmed that the organising SC--FC features that we here identify are not crucially dependent on the binarisation, thresholding and normalisation procedure, described in \textbf{\textit{Structural and functional connectivity}} and are qualitatively similar under variation of coupling strength (see \textbf{MATHEMATICAL METHODS}); moreover, results obtained via MPC and of MPA are indistinguishable (data not shown).} In Fig.~\ref{fig:results}(b) we show a plot of $H'(0)$. Recall from \textbf{\textit{Weakly-coupled oscillator theory}} that a globally synchronous state (which is guaranteed to exist from the row-sum constraint) is stable if $\varepsilon H'(0)>0$. Comparison with Fig. \ref{fig:results}(a), highlights that when synchrony is unstable ($\varepsilon H'(0)< 0$) SC only weakly drives FC. \Red{Moreover, this instability region coincides with the region of bistability and the false bifurcation, stressing the important role of these bifurcations for understanding SC--FC correlation.}

Of course, there is a much finer structure in Fig.~\ref{fig:results}(a) that is not predicted by considering either the bifurcation from steady state, or the weakly-coupled analysis of synchronous states, and so it is illuminating to pursue the full weakly coupled oscillator analysis for structured networks. The eigenvalues of the Jacobian, corresponding to more general stable phase-locked states, can be used to give a measure of solution attractivity.  The largest eigenvalue is plotted in Fig.~\ref{fig:results}(c). \Red{The most stable (non-synchronous) phase-locked states occur in the neighbourhood of the false bifurcations, as well as in the region of bistability and along the existence border for oscillations, defined by a saddle node bifurcation.} Furthermore, apart from near false bifurcations, stronger stability of the general phase-locked states corresponds with stronger stability of global synchrony (Fig. \ref{fig:results}(b)).

\begin{figure}[!ht]
	\centering
	\includegraphics[width=\textwidth]{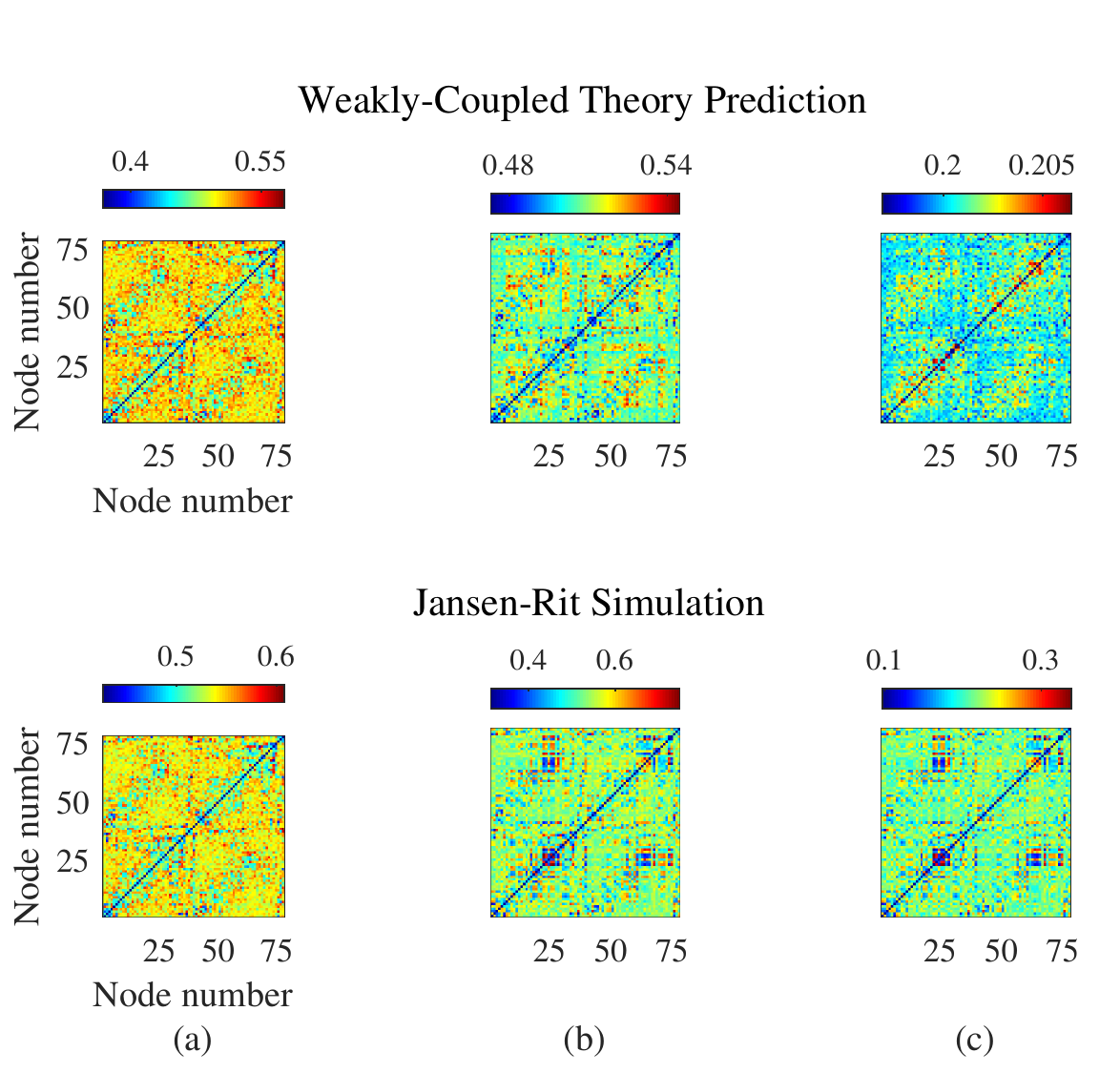}
	\caption{Comparison of FC patterns from averages of realisations of the weakly-coupled oscillator model \Red{(\ref{eq:phasedynamics})} with corresponding Jansen--Rit \Red{(\ref{eq:JR})} simulations, \RedText{with no noise present}, at $A=5, B=19$, computing averages over 600 realisations with initial conditions chosen at random (other parameter values are given in Table \ref{tab:JR}). (a) $\varepsilon$=0.01; (b) $\varepsilon$=0.1; (c) $\varepsilon$=1. These results show how the weakly-coupled theory becomes less predictive for stronger coupling strengths, resulting in matrices with Jaccard similarity of 0.98, 0.76 and 0.65 (to 2 s.f.) respectively.}
	\label{fig:FCplots}
\end{figure}

\begin{figure}[!ht]
\centering
		\includegraphics[width=\textwidth]{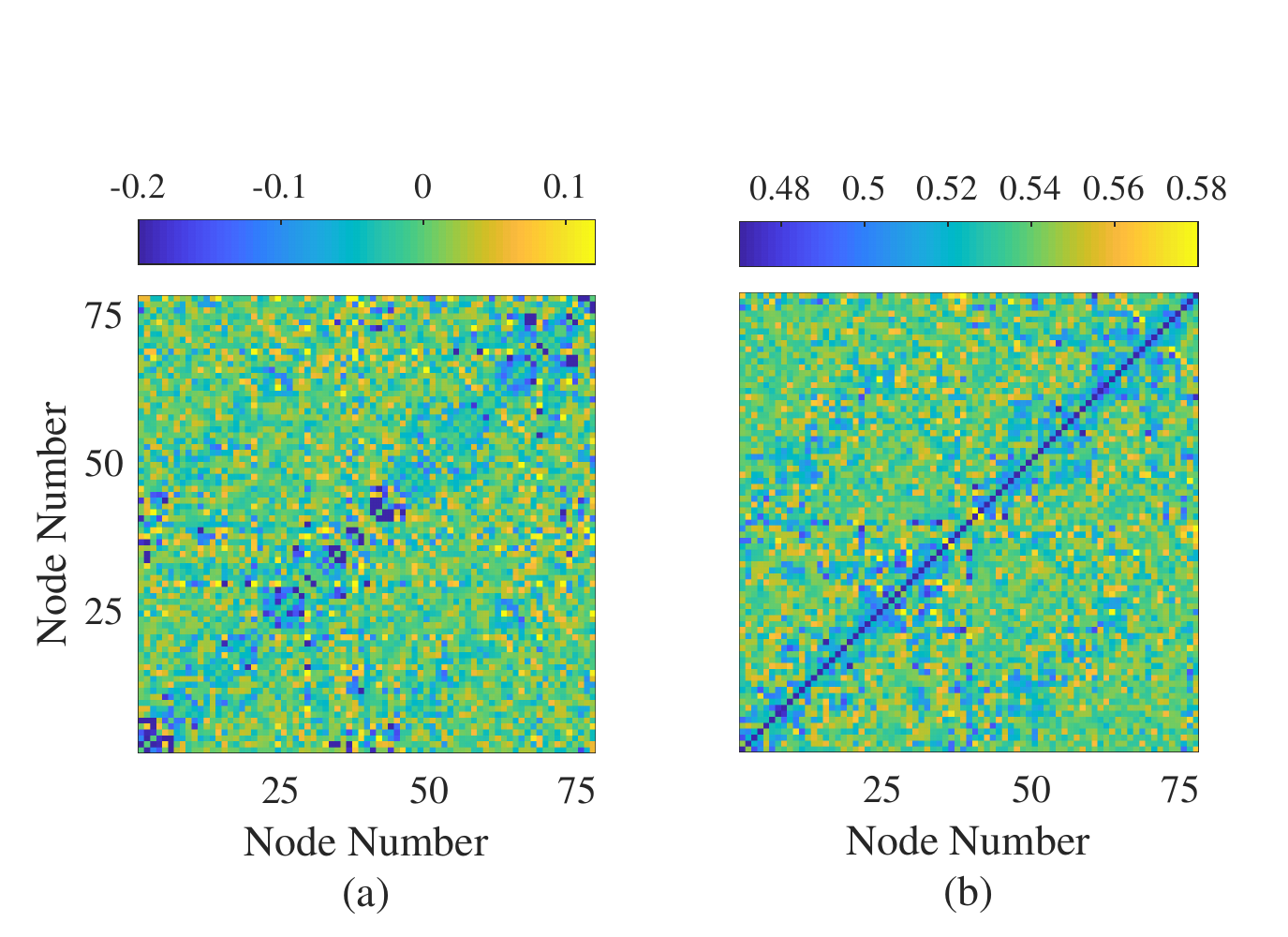}
	\caption{\Red{(a) FC prediction given by the a linear combination of eigenmodes of the weakly-coupled oscillator system, given by tensor products of eigenvectors of the SC graph Laplacian (\ref{eq:Rjk_predicted}), with $N^*=N$. (b) Direct simulation of the Jansen--Rit network model (\ref{eq:JR}) \RedText{with no noise present}. Parameter values are chosen as $A=6$, $B=18$, which lies near the existence border for stable synchronous solutions (see Fig.~\ref{fig:results}(b)); other parameter values are given in Table \ref{tab:JR}. The (weighted) Jaccard similarity between the two FC networks (scaled to $[0,1]$ for comparability) is calculated to be 0.82, indicating the predictive power of equation (\ref{eq:Rjk_predicted}).}}
\label{fig:eig}
\end{figure}

To test the predictive power of the weakly-coupled theory, in Fig.~\ref{fig:FCplots} we compare the emergent FC structure obtained from direct simulations of the Jansen--Rit network model (\ref{eq:JR}) against direct simulations of the weakly-coupled oscillator network \eqref{eq:phasedynamics}. For the former, the phases required to compute the mean phase agreement (equation \eqref{eq:MPA}) are determined from each timeseries by a Hilbert transform; in the latter case, the phase variables from equation (\ref{eq:phasedynamics}) are employed directly. \RedText{Since the weakly-coupled reduction of the Jansen--Rit model is deterministic, these computations were ran in the absence of noise ($dW_i=0$ for all nodes).} \Red{As expected, we find excellent agreement between the modular FC structure in the case for very weak coupling, with this agreement reducing with increasing $\varepsilon$, as quantified by a reduction in Jaccard similarity (from 0.98 in panel (a) to 0.65 in (c)). This is a manifestation of the network moving from a dynamical regime that can be well described by the weakly-coupled reduction \eqref{eq:phasedynamics} to one where stronger network interactions dominate. Since an analogous theory does not exist for stronger coupling, we do not consider here how SC--FC relations arise from network dynamics within a strongly--coupled framework.}  \Red{Moreover, through the instability theory of the synchronous state we can construct a proxy for the FC as described in \textbf{\textit{Weakly-coupled oscillator theory}}. In Fig.~\ref{fig:eig} we compare simulated FC with that predicted by $\hat{R}$ (equation (\ref{eq:Rjk_predicted}); \textit{i.e.}~using the unstable eigenmodes of the Jacobian at synchrony), for parameter values that lie just beyond the onset of instability of the globally synchronous state and near the false bifurcation set (see Figs~\ref{fig:results}(a,b)). We observe that the key features of the FC are captured by the eigenmode prediction; indeed the (weighted) Jaccard similarity coefficient between predicted and simulated FC (both scaled to $[0,1]$) is calculated to be 0.82. This is a much more efficient way of simulating an emergent FC pattern, since it does not require brute-force forward integrations of the model, which may take a long time to converge.}

\Red{All of these results highlight the strong impact that nodal dynamics can have on the correlation between SC and FC, and the utility of bifurcation theory and phase oscillator reduction techniques (that are naturally positioned  to  explain  the  generation  of  patterns  of  synchronous  node  and  network  activity)  to provide insight into how SC–-FC correlations are organised across parameter space.}
   
\section{Discussion}\label{sec:Discussion}

In this paper, we investigate the degree to which the dynamical state of neural populations, as well as their structural connectivity, facilitates the emergence of functional connections in a neural-mass network model of the human brain. We have addressed this by using a mixture of computational and mathematical techniques to assess the correlation between structural and functional connectivity as one traverses the parameter space controlling the inhibitory and excitatory dynamics and bifurcations of an isolated Jansen--Rit neural mass model. Importantly, SC has been estimated from HCP diffusion MRI datasets. We find that SC strongly drives FC when the system is close to a Hopf bifurcation, whereas in the neighbourhood of a false bifurcation, this drive is diminished. These results emphasise the vital role that local dynamics has to play in determining FC in a network with a static SC. In addition, we show that a weakly-coupled analysis provides insight into the organisation of SC--FC correlation features across parameter space, and can be exploited to predict emergent FC structure. \Red{\citet{messe2014relating} considered statistical models to predict FC from SC (in particular, a spatial simultaneous autoregressive model (sSAR), whose parameters can be estimated in a Bayesian framework) and found, interestingly, that simpler linear models were able to fare at least as well. More recently, \citet{saggio2016analytical} were also able to make predictions of FC from empirical SC data (and vice versa) using a simple linear model. Since the only free parameter of their model for SC is the global coupling strength, results from this method are efficient and computationally inexpensive. We have not attempted to reproduce empirical data here, but we have show that similar predictions can be made using bifurcation theory and network reduction techniques; such an approach allows us to consider in more detail, and explain, the influence of the rich neural dynamics supported by the Jansen--Rit model on SC--FC relationships. Nevertheless, it is important to note that the FC structures we are concerned with are averaged over long-time scales and therefore represent a static FC state, as opposed to dynamic FC (as discussed in \textbf{INTRODUCTION}). Use of such static FC networks as a clinical biomarker is widespread; however, subject variability in FC means that their predictive power is restricted to group analyses \citep{mueller2013individual}. To capture the rich dynamic FC repertoire exhibited in empirical resting state data, for example the distinct hierarchical organisation in switching between FC states \citep{vidaurre2017brain}, will require alternative approaches. One such approach is dynamic causal modelling, as employed in \citet{goulden2014salience} and \citep{van2019dynamic} for empirical data.}

The modelling work presented here is relevant in a wider neuroimaging context---for example, epilepsy is often considered to be caused by irregularities in synchronisation \citep{mormann2003epileptic,netoff2002decreased,lehnertz2009synchronization}. It is noteworthy that the changes in synchrony patterns that we observe arise from local dynamical considerations as opposed to large scale structural ones. In the Jansen--Rit model, the bifurcations organising emergent FC take the form of Hopf, saddle, \Red{fold of limit cycle} and false bifurcations. False bifurcations have received relatively little attention in the dynamical systems community (a notable exception being the work of \citet{Marten2009}), although our results indicate that they may be significant for understanding how `synchronisability' of brain networks is reduced during seizures. This phenomena was reported in \citet{schindler2008evolving}, which also found that synchronisability increases as the patient recovers from seizure state.

\Red{A natural extension to the work presented here would be the inclusion of conduction delays, characterised by Euclidean or path-length distances between brain regions, which are certainly important in modulating the spatiotemperal coherence in the brain \citep{deco2009key}. These would manifest as constant phase shifts in the weakly-coupled reduction of the model \citep{ton2014structure}.  For strongly coupled systems the mathematical treatment of networks with delayed interactions remains an open challenge. Recent work in this vein by \citet{tewarie2019spatially} focusses on the role of delays in destabilising network steady states, and techniques extending the Master Stability Function to delayed systems \citep{otto2018synchronization} may be appropriate for treating phase-locked network states.}

In summary, the findings reported here suggest that there are multiple factors which give rise to emergent FC. While structure clearly facilitates functional connectivity, the degree to which it influences emergent FC states is determined by the dynamics of its neural sub-units. Importantly, we have shown that local dynamics has a clear influence on SC--FC correlation, as does network topology and coupling strength. Our combined mathematical and computational study has demonstrated that a full description of the mechanisms that dictate the formation of FC from anatomy requires knowledge of how both neuronal activity and connectivity are modulated and, moreover, exposes the utility of bifurcation theory and network reduction techniques. This work can be extended to more complex neural mass models such as that derived in \citet{Coombes2019}, to further explore the relationship between dynamics and structure--function relations in systems with more sophisticated models for node dynamics.

\section{Mathematical Methods}

\subsection{Network bifurcations of equilibria}
\label{appendixA}

Each node of the network is described by the $m$-dimensional Jansen--Rit model, with $m=6$.  There are $N=78$ nodes in the network.
Analysing bifurcations of network equilibria requires finding a set of $m \times N$ eigenvalues from the linearised system.
Defining $\mathbf{y}_i \in \RSet^m$ as $(y_{1_i},\dots ,y_{m_i})^\intercal$ allows us to write the system of first-order ODEs given by \eqref{eq:JR} for the network as:
\begin{equation}
\FD{}{t}\mathbf{y}_i=M \mathbf{y}_i+B+L f\left(\tau\mathbf{y}_i\right)+ \varepsilon aA \sum_{j=1}^N w_{ij} \mathbf{K} (\mathbf{y}_j ) , \qquad i=1,\ldots, N,
\label{eq:MSF}
\end{equation}
where
\begin{equation}
M=\begin{bmatrix}
0_3 & I_3 \\
M_{21} & M_{22}
\end{bmatrix}, \qquad
L = \begin{bmatrix}
0_3 & 0_3 \\
0_3 & L_{22}
\end{bmatrix} , \qquad
\tau = 
\begin{bmatrix}
0_3 & 0_3 \\
\tau_{21} & 0_3
\end{bmatrix},
\end{equation}
and $B =(0,0,0,0,AaP,0)^\intercal$, $\mathbf{K}(\mathbf{y}) = (0,0,0,0,f(y_1-y_2),0)^\intercal$, with
\begin{align}
M_{21} &= -\begin{bmatrix}
a^2 & 0 & 0\\
0 & a^2 & 0 \\
0 & 0 & b^2
\end{bmatrix}, \qquad
M_{22} = - 2 \begin{bmatrix}
a & 0 & 0\\
0 & a & 0 \\
0 & 0 & b
\end{bmatrix}, \\
L_{22} &= \begin{bmatrix}
Aa & 0 & 0\\
0 & Aa C_2 & 0 \\
0 & 0 & Bb C_4
\end{bmatrix}, \qquad
\tau_{21} = \begin{bmatrix}
0 & 1 & -1\\
C_1 & 0 & 0 \\
C_3 & 0 & 0
\end{bmatrix}.
\end{align}
Here we have introduced the $3 \times 3$ identity matrix $I_3$, and the $3 \times 3$ zero matrix $0_3$.  The network steady state $\mathbf{y}_i = \overline{\mathbf{y}}_i$, for $i=1,\ldots,N$, is defined by setting the left hand side of (\ref{eq:MSF}) to zero.  We now linearise \eqref{eq:MSF} by setting $\mathbf{y}_i(t)=\overline{\mathbf{y}}_i+\mathbf{u}_i(t)$, where $\mathbf{u}_i(t)$ is a small perturbation.  
This gives,
\begin{equation}
\FD{}{t}\mathbf{u}_i= \left [M  + Lf' \left(\tau \overline{\mathbf{y}}_i\right) \tau \right ] \mathbf{u}_i + \varepsilon aA \sum_{j=1}^N w_{ij} D\mathbf{K} (\overline{\mathbf{y}}_j ) \mathbf{u}_j , 
\label{uidot}
\end{equation}
where $D \mathbf{K}(\mathbf{y}) \in \RSet^{m \times m}$ is the Jacobian of $\mathbf{K}(\mathbf{y})$.  The only two non-zero entries of this matrix are given by $[\mathbf{K}(\mathbf{y})]_{5,2} = f'(\overline{y}_1-\overline{y}_2)  = -[\mathbf{K}(\mathbf{y})]_{5,3}$.
It is now useful to define $D\mathbf{F}_i\equiv\left[M  + Lf' \left(\tau \overline{\mathbf{y}}_i\right) \tau \right]$ and $D\mathbf{G}_j\equiv \varepsilon aAD\mathbf{K} (\overline{\mathbf{y}}_j)$,
so that $D\mathbf{F}_i$ is the Jacobian which describes the intra-mass dynamics of node $i$ and $D\mathbf{G}_j$ is the Jabobian for the effect of the inter-mass interactions with node $j$. Then we may write (\ref{uidot}) in the form
\begin{equation}
\FD{}{t} \mathbf{U} = \begin{bmatrix}
D \mathbf{F}_1 &  & 0\\
& \ddots & \\
0 & & D \mathbf{F}_N
\end{bmatrix} \mathbf{U} + 
( w \otimes I_m)
\begin{bmatrix}
D \mathbf{G}_1 &  & 0\\
& \ddots & \\
0 & & D \mathbf{G}_N
\end{bmatrix} \mathbf{U} ,
\label{Udot}
\end{equation}
where $U=(\mathbf{u}_1, \ldots, \mathbf{u}_N)^\intercal$, and $\otimes$ denotes the tensor product.
This system can be simplified by considering the eigenvalues of the connectivity matrix $w \in \RSet^{N \times N}$ (with components $w_{ij}$). We introduce a matrix of normalised eigenvectors, $E$, and a corresponding diagonal matrix of eigenvalues, $\Lambda=\operatorname{diag} (\mu_1 \dots \mu_{N})$, such that $w E=E \Lambda$. Imposing the change of variables $\mathbf{V}=(E \otimes I_m)^{-1}\mathbf{U}$ transforms (\ref{Udot}) to
\begin{align}
\FD{}{t} \mathbf{V} &= (E \otimes I_m)^{-1} \begin{bmatrix}
D \mathbf{F}_1 &  & 0\\
& \ddots & \\
0 & & D \mathbf{F}_N
\end{bmatrix} (E \otimes I_m) \mathbf{V} \nonumber \\
&+ (E \otimes I_m)^{-1}
( w \otimes I_m)
\begin{bmatrix}
D \mathbf{G}_1 &  & 0\\
& \ddots & \\
0 & & D \mathbf{G}_N
\end{bmatrix} (E \otimes I_m) \mathbf{V}
\label{Vdot} .
\end{align}
Assuming a homogeneous system such that $\bar{x}_i$ is independent of $i$, which is natural for identical units with a network connectivity with a row-sum constraint $\sum_{j=1}^{N}w_{ij}=\mbox{const}$ for all $i$, then we have a useful simplification $D\mathbf{F}_i=D\mathbf{F}$ and $D\mathbf{G}_i=D\mathbf{G}$ for all $i$. It is simple to establish that for any block diagonal matrix $A$, formed from $N$ equal matrices of size $m \times m$, that
$(E \otimes I_m)^{-1} A (E \otimes I_m)=A$.  Moreover, using standard properties of the tensor operator, $(E \otimes I_m)^{-1}( w \otimes I_m) = (E^{-1} w) \otimes I_m=(\Lambda E^{-1}) \otimes I_m = (\Lambda \otimes I_m)(E^{-1} \otimes I_m)$.  Hence, (\ref{Vdot}) becomes
\begin{equation}
\FD{}{t} \mathbf{V} = \begin{bmatrix}
D \mathbf{F} &  & 0\\
& \ddots & \\
0 & & D \mathbf{F}
\end{bmatrix} \mathbf{V} + 
\begin{bmatrix}
\mu_1 D \mathbf{G} &  & 0\\
& \ddots & \\
0 & & \mu_N D \mathbf{G}
\end{bmatrix} \mathbf{V}.
\label{Vdot1}
\end{equation}
The system (\ref{Vdot1}) is in a block diagonal form and so it is equivalent to the set of decoupled equations given by
\begin{equation}
\FD{}{t} \xi_p = \left [D \mathbf{F} + \mu_p D \mathbf{G} \right ] \xi_p, \qquad \xi_p \in \CSet^m , \qquad p=1,\dots, N .
\end{equation}
This has solutions of the form $\xi_p = \mathbf{A}_p \e^{\lambda t}$ for some amplitude vector $\mathbf{A}_p \in \CSet^m$.  For a non-trivial set of solutions we require $\mathcal{E}(\lambda;p)=0$ where
\begin{equation}
\mathcal{E}(\lambda;p) = \det \left [\lambda I_m - D \mathbf{F} - \mu_p D \mathbf{G} \right ] ,\qquad p=1,\ldots,N .
\label{spectral}
\end{equation} 
Solving $\mathcal{E}=0$ for $\lambda$ produces a set of eigenvalues which can be tracked to determine bifurcations. Since \RedText{local stability} requires the real part of all eigenvalues to be negative, if one of these eigenvalues crosses the imaginary axis the solution can undergo either a saddle-node bifurcation ($\text{Re} \, \lambda=0=\text{Im}\,(\lambda)$) or a Hopf bifurcation ($\text{Re} \, \lambda=0$, $\text{Im}\,(\lambda) \neq 0$).

\subsection{Phase interaction function}
\label{appendixB}

To investigate the nature of phase-locked oscillatory states in the Jansen--Rit network, it is appropriate to use weakly-coupled oscillator theory.  For a recent review see \citep{ashwin2016mathematical}.  This gives rise to the set of equations (\ref{eq:phasedynamics}) where the phase interaction function $H$ is determined in terms of two quantities.  The first is the so-called phase response or adjoint $\mathbf{Q} \in \RSet^m$, that describes the response of an attracting limit cycle to a small perturbation.  This can be computed by solving the \textit{adjoint equation}.  It is convenient to write the dynamics for a single uncoupled Jansen--Rit node in the form $\dot{\mathbf{y}}= \mathbf{F}(\mathbf{y})$, with $\mathbf{F}, \mathbf{y} \in \RSet^m$. Using the notation above we have 
explicitly that $\mathbf{F}(\mathbf{y}) = M \mathbf{y}+B+Lf\left(\tau\mathbf{y}\right)$. The adjoint
is given by the $T$-periodic solution of
\begin{equation}
\FD{}{t} \mathbf{Q} = - D \mathbf{F}^\intercal(\overline{\mathbf{y}}(t)) \mathbf{Q}, \qquad \langle \mathbf{Q}(0), \mathbf{F}(\overline{\mathbf{y}}(0)) = \Omega .
\end{equation}
Here $\overline{\mathbf{y}}(t)$ is a $T$-periodic of the Jansen--Rit node model and $\langle~,~ \rangle$ denotes a Euclidean inner product between vectors.  The second ingredient comes from writing the physical interactions in terms of phases rather than the original state variables.  This is easily done by writing $\mathbf{y}_i(t) = \overline{\mathbf{y}}(\theta_i/\Omega)$.  The phase interaction function is then obtained as
\begin{equation}
	H(t) = \frac{1}{T}\int_0^T \d s  \big\langle \mathbf{Q}(s), aA \mathbf{K}(\overline{\mathbf{y}}(s+t))
\label{H}
\big\rangle .
\end{equation}
The adjoint equation is readily solved numerically by backward integration in time \citep{Williams1997}, whilst the integral in (\ref{H}) can be evaluated using numerical quadrature.

\subsection{Structural connectivity data}
\label{appendixC}

\begin{figure}[!ht]
\centering
		\includegraphics[width=0.7\textwidth]{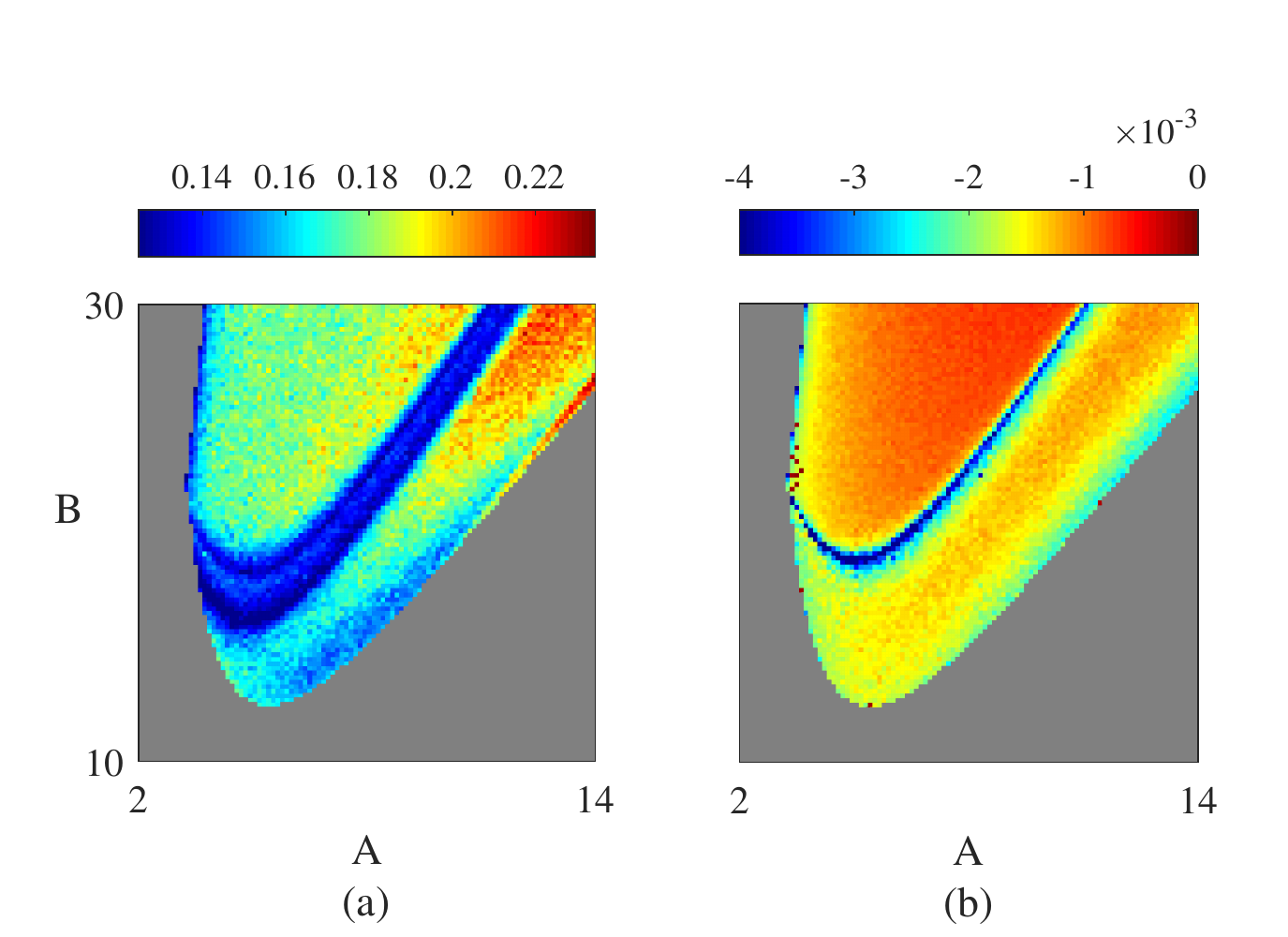}
	\caption{(a) Jaccard similarity coefficient between SC and FC in numerical simulations of the Jansen--Rit network model (\ref{eq:JR}), when the network supports an oscillatory solution. Here the structural connectivity is the original weighted, un-normalised data. Model parameters are as in Fig.~\ref{fig:results}. (b) The largest non-zero eigenvalue of the Jacobian for the full weakly-coupled oscillator network (equation (5)), calculated at a stable phase-locked state for the un-normalised SC matrix.}
\label{fig:Jaccard_unnormalised}
\end{figure}

\Red{As described in \textit{\textbf{Structural and functional connectivity}}, we process structural connectivity data obtained from the HCP by thresholding, binarising and normalising by row. To confirm that these procedures do not unduly influence our conclusions, or restrict their applicability, we performed the following tests. 

Statistical checks on the distribution of unthresholded SC weights indicate that node degree distributions have standard deviation of less than 10\% of the mean, and outliers differ from the mean by less than $25\%$ (data omitted). Therefore we are confident that our thresholding and binarisation process does not unduly influence the SC network structure, and thereby our results. As noted in the main text, we have also confirmed that the features of SC--FC correlation that we uncover in Fig.~\ref{fig:results}(a) are retained for different thresholds (namely: 20\%, 30\%, 40\%; data not shown). To ensure that our modifications to the SC matrix did not crucially influence our findings, we recalculate equivalents of Figures \ref{fig:results}(a) and (c) for a weighted, unnormalised network, obtaining similar SC--FC structures (see Figure \ref{fig:Jaccard_unnormalised}). Inspection of node behaviour in the weighted un-normalised network, at parameter choices for which Figure 5(b) predicts stable or unstable synchronous behaviour, shows that the predictive power of our linear analysis is retained in the unnormalised case (data not shown).}

\begin{figure}[!ht]
	\centering
		\includegraphics[width=\textwidth]{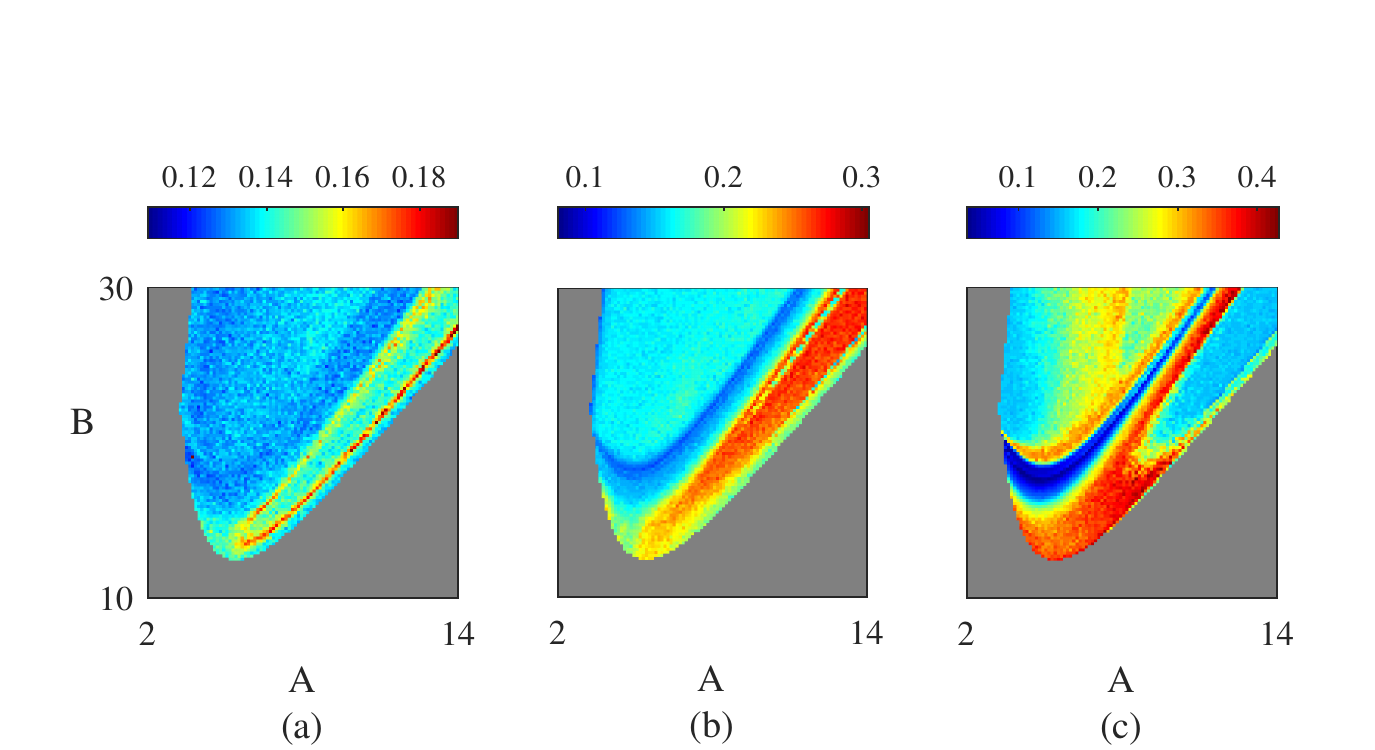}
		\caption{Jaccard similarity coefficient between SC and FC in numerical simulations of the Jansen--Rit network model (\ref{eq:JR}), when the network supports an oscillatory solution. Parameters are as in Fig.~\ref{fig:results}, except (a) $\varepsilon=0.01$, (b) $\varepsilon=0.1$, (c) $\varepsilon=1.0$. Warmer colours indicate greater SC--FC correlation.}\label{fig:jaccardcoupling}
\end{figure}

\Red{As noted in \citet{hansen2015functional}, variation in coupling strength can affect SC--FC relations. In Fig.~\ref{fig:jaccardcoupling}, we show that the essential organising features of the Jaccard similarity between SC and FC that we highlight in Fig.~\ref{fig:results}(a) are qualitatively unchanged for a range of choices of coupling strength $\varepsilon$}.

\section{Acknowledgements}

This work was supported by the Engineering and Physical Sciences Research Council [grant number EP/N50970X/1].



\bibliography{JansenRit}   

\newpage
\section*{A list of Technical Terms}

\begin{description}
\item[Structural connectivity:] A pattern of anatomical links between between distinct brain regions.  This is often measured using diffusion tensor imaging.
\item[Diffusion tensor imaging:]  A magnetic resonance imaging technique that measures the diffusion of water in tissue in order to produce axonal fibre tract images.
\item[Functional connectivity:] A pattern of statistical dependencies between between distinct brain regions.  This is often measured using coherence or correlation measures between time-series.
\item[Human Connectome Project:] A consortium research project to build a human connectome for structural and functional connectivity, and facilitate research into brain disorders.
\item[Neural mass model:] A phenomenological model for the activity of a neuronal population cast as a system of ordinary differential equations.
\item[Jaccard similarity:]  An index for comparing members for two sets of binary data to see which are shared and which are distinct. The higher the index, the more similar the two sets.
\item[Hopf bifurcation:]  The appearance or disappearance of a periodic orbit through a local change in the stability properties of an equilibrium point under parameter variation. 
\item[Saddle-node bifurcation:] A local bifurcation in which two equilibria of a dynamical system collide and annihilate.
\item[False bifurcation:]  A qualitative change in the shape of a periodic orbit without a change in stability, say from a small to a large amplitude oscillation, that occurs under a very small parameter variation.
\item[Weakly coupled oscillator theory:] A reduction of a network of weakly interacting limit cycle oscillators to a system of fewer dynamical equations that describes the evolution of relative phases between nodes.
\end{description}

\end{document}